\documentclass[%
 reprint,
superscriptaddress,
preprintnumbers,
 amsmath,amssymb,
 aps,
showpacs]{revtex4-1}

\usepackage{graphicx}% Include figure files
\usepackage{dcolumn}% Align table columns on decimal point
\usepackage{bm}% bold math
\usepackage{amssymb}
\usepackage{amsmath}
\usepackage{hyperref}% add hypertext capabilities
\usepackage{units}
\usepackage{xspace}
\usepackage{multirow}
\usepackage{xcolor}
\usepackage[tight]{subfigure}
\usepackage[mathlines]{lineno}% Enable numbering of text and display math
\relax 

\newcommand{\nue}{\ensuremath{\nu_{e}}\xspace}
\newcommand{\nuebar}{\ensuremath{\overline{\nu}_{e}}\xspace}
\newcommand{\numubar}{\ensuremath{\overline{\nu}_{\mu}}\xspace}
\newcommand{\numu}{\ensuremath{\nu_{\mu}}\xspace}
\newcommand{\nutau}{\ensuremath{\nu_{\tau}}\xspace}
\newcommand{\nutaubar}{\ensuremath{\overline{\nu}_{\tau}}\xspace}

\begin{document}
\preprint{FERMILAB-PUB-20-253-ND}
\title{Precision constraints for three-flavor neutrino oscillations from the full MINOS+ and MINOS data set}
\newcommand{\Berkeley}{Lawrence Berkeley National Laboratory, Berkeley, California, 94720 USA}
\newcommand{\Cambridge}{Cavendish Laboratory, University of Cambridge, %Madingley Road, 
Cambridge CB3 0HE, United Kingdom}
\newcommand{\Cincinnati}{Department of Physics, University of Cincinnati, Cincinnati, Ohio 45221, USA}
\newcommand{\FNAL}{Fermi National Accelerator Laboratory, Batavia, Illinois 60510, USA}
\newcommand{\RAL}{Rutherford Appleton Laboratory, Science and Technology Facilities Council, Didcot, OX11 0QX, United Kingdom}
\newcommand{\UCL}{Department of Physics and Astronomy, University College London, 
%Gower Street, 
London WC1E 6BT, United Kingdom}
\newcommand{\Caltech}{Lauritsen Laboratory, California Institute of Technology, Pasadena, California 91125, USA}
\newcommand{\Alabama}{Department of Physics and Astronomy, University of Alabama, Tuscaloosa, Alabama 35487, USA}
\newcommand{\ANL}{Argonne National Laboratory, Argonne, Illinois 60439, USA}
\newcommand{\Athens}{Department of Physics, University of Athens, GR-15771 Athens, Greece}
\newcommand{\NTUAthens}{Department of Physics, National Tech. University of Athens, GR-15780 Athens, Greece}
\newcommand{\Benedictine}{Physics Department, Benedictine University, Lisle, Illinois 60532, USA}
\newcommand{\BNL}{Brookhaven National Laboratory, Upton, New York 11973, USA}
\newcommand{\CdF}{APC -- Universit\'{e} Paris 7 Denis Diderot, 10, rue Alice Domon et L\'{e}onie Duquet, F-75205 Paris Cedex 13, France}
\newcommand{\Cleveland}{Cleveland Clinic, Cleveland, Ohio 44195, USA}
\newcommand{\Delhi}{Department of Physics \& Astrophysics, University of Delhi, Delhi 110007, India}
\newcommand{\GEHealth}{GE Healthcare, Florence South Carolina 29501, USA}
\newcommand{\Harvard}{Department of Physics, Harvard University, Cambridge, Massachusetts 02138, USA}
\newcommand{\HolyCross}{Holy Cross College, Notre Dame, Indiana 46556, USA}
\newcommand{\Houston}{Department of Physics, University of Houston, Houston, Texas 77204, USA}
\newcommand{\IIT}{Department of Physics, Illinois Institute of Technology, Chicago, Illinois 60616, USA}
\newcommand{\Iowa}{Department of Physics and Astronomy, Iowa State University, Ames, Iowa 50011 USA}
\newcommand{\Indiana}{Indiana University, Bloomington, Indiana 47405, USA}
\newcommand{\ITEP}{High Energy Experimental Physics Department, ITEP, B. Cheremushkinskaya, 25, 117218 Moscow, Russia}
\newcommand{\JMU}{Physics Department, James Madison University, Harrisonburg, Virginia 22807, USA}
\newcommand{\LANL}{Los Alamos National Laboratory, Los Alamos, New Mexico 87545, USA}
\newcommand{\Lebedev}{Nuclear Physics Department, Lebedev Physical Institute, Leninsky Prospect 53, 119991 Moscow, Russia}
\newcommand{\Lancaster}{Lancaster University, Lancaster, LA1 4YB, UK}
\newcommand{\LLL}{Lawrence Livermore National Laboratory, Livermore, California 94550, USA}
\newcommand{\LosAlamos}{Los Alamos National Laboratory, Los Alamos, New Mexico 87545, USA}
\newcommand{\Manchester}{Department of Physics and Astronomy, University of Manchester, 
%Oxford Road, 
Manchester M13 9PL, United Kingdom}
\newcommand{\MIT}{Lincoln Laboratory, Massachusetts Institute of Technology, Lexington, Massachusetts 02420, USA}
\newcommand{\Minnesota}{University of Minnesota, Minneapolis, Minnesota 55455, USA}
\newcommand{\Crookston}{Math, Science and Technology Department, University of Minnesota -- Crookston, Crookston, Minnesota 56716, USA}
\newcommand{\Duluth}{Department of Physics, University of Minnesota Duluth, Duluth, Minnesota 55812, USA}
\newcommand{\Ohio}{Center for Cosmology and Astro Particle Physics, Ohio State University, Columbus, Ohio 43210 USA}
\newcommand{\Otterbein}{Otterbein University, Westerville, Ohio 43081, USA}
\newcommand{\Oxford}{Subdepartment of Particle Physics, University of Oxford, Oxford OX1 3RH, United Kingdom}
\newcommand{\PennState}{Department of Physics, Pennsylvania State University, State College, Pennsylvania 16802, USA}
\newcommand{\PennU}{Department of Physics and Astronomy, University of Pennsylvania, Philadelphia, Pennsylvania 19104, USA}
\newcommand{\Pittsburgh}{Department of Physics and Astronomy, University of Pittsburgh, Pittsburgh, Pennsylvania 15260, USA}
\newcommand{\IHEP}{Institute for High Energy Physics, Protvino, Moscow Region RU-140284, Russia}
\newcommand{\Rochester}{Department of Physics and Astronomy, University of Rochester, New York 14627 USA}
\newcommand{\RoyalH}{Physics Department, Royal Holloway, University of London, Egham, Surrey, TW20 0EX, United Kingdom}
\newcommand{\Carolina}{Department of Physics and Astronomy, University of South Carolina, Columbia, South Carolina 29208, USA}
\newcommand{\SDakota}{South Dakota School of Mines and Technology, Rapid City, South Dakota 57701, USA}
\newcommand{\SLAC}{Stanford Linear Accelerator Center, Stanford, California 94309, USA}
\newcommand{\Stanford}{Department of Physics, Stanford University, Stanford, California 94305, USA}
\newcommand{\StJohnFisher}{Physics Department, St. John Fisher College, Rochester, New York 14618 USA}
\newcommand{\Sussex}{Department of Physics and Astronomy, University of Sussex, Falmer, Brighton BN1 9QH, United Kingdom}
\newcommand{\TexasAM}{Physics Department, Texas A\&M University, College Station, Texas 77843, USA}
\newcommand{\Texas}{Department of Physics, University of Texas at Austin, 
%1 University Station C1600, 
Austin, Texas 78712, USA}
\newcommand{\TechX}{Tech-X Corporation, Boulder, Colorado 80303, USA}
\newcommand{\Tufts}{Physics Department, Tufts University, Medford, Massachusetts 02155, USA}
\newcommand{\UNICAMP}{Universidade Estadual de Campinas, IFGW, CP 6165, 13083-970, Campinas, SP, Brazil}
\newcommand{\UFG}{Instituto de F\'{i}sica, 
Universidade Federal de Goi\'{a}s, 74690-900, Goi\^{a}nia, GO, Brazil}
\newcommand{\USP}{Instituto de F\'{i}sica, Universidade de S\~{a}o Paulo,  CP 66318, 05315-970, S\~{a}o Paulo, SP, Brazil}
\newcommand{\Warsaw}{Department of Physics, University of Warsaw, %Pasteura 5, 
PL-02-093 Warsaw, Poland}
\newcommand{\Washington}{Physics Department, Western Washington University, Bellingham, Washington 98225, USA}
\newcommand{\WandM}{Department of Physics, College of William \& Mary, Williamsburg, Virginia 23187, USA}
\newcommand{\Wisconsin}{Physics Department, University of Wisconsin, Madison, Wisconsin 53706, USA}
\newcommand{\deceased}{Deceased.}
\newcommand{\Dallas}{Department of Physics, University of Dallas, Irving, Texas 75062, USA }

%\affiliation{\ANL}
%\affiliation{\Athens}
%\affiliation{\Benedictine}
%\affiliation{\BNL}
%\affiliation{\Caltech}
\affiliation{\Cambridge}
%\affiliation{\UNICAMP}
%\affiliation{\CdF}
\affiliation{\Cincinnati}
\affiliation{\Dallas}
\affiliation{\FNAL}
\affiliation{\UFG}
\affiliation{\Harvard}
%\affiliation{\HolyCross}
\affiliation{\Houston}
%\affiliation{\IIT}
%\affiliation{\Indiana}
\affiliation{\Iowa}
%\affiliation{\IHEP}
%\affiliation{\ITEP}
%\affiliation{\JMU}
%\affiliation{\Lebedev}
%\affiliation{\LLL}
\affiliation{\Lancaster}
\affiliation{\UCL}
\affiliation{\Manchester}
\affiliation{\Minnesota}
\affiliation{\Duluth}
%\affiliation{\Otterbein}
\affiliation{\Oxford}
\affiliation{\Pittsburgh}
\affiliation{\RAL}
\affiliation{\USP}
%\affiliation{\Carolina}
\affiliation{\Stanford}
\affiliation{\Sussex}
%\affiliation{\TexasAM}
\affiliation{\Texas}
\affiliation{\Tufts}
\affiliation{\Warsaw}
%\affiliation{\Washington}
\affiliation{\WandM}
\affiliation{\Wisconsin}

\author{P.~Adamson}
\affiliation{\FNAL}
%\affiliation{\UCL}
%\affiliation{\Sussex}

%\author{C.~Andreopoulos}
%\affiliation{\RAL}
%\affiliation{\Athens}

\author{I.~Anghel}
\affiliation{\Iowa}
%\affiliation{\ANL}

%\author{K.~E.~Arms}
%\affiliation{\Minnesota}

%\author{R.~Armstrong}
%\affiliation{\Indiana}

\author{A.~Aurisano}
\affiliation{\Cincinnati}

%\author{T.~H.~Fields}
%\affiliation{\ANL}

%\author{D.~J.~Auty}
%\affiliation{\Sussex}

%\author{S.~Avvakumov}
%\affiliation{\Stanford}

%\author{D.~S.~Ayres}
%\affiliation{\ANL}

%\author{C.~Backhouse}
%\affiliation{\Oxford}

%\author{B.~Baller}
%\affiliation{\FNAL}

%\author{B.~Barish}
%\affiliation{\Caltech}

%\author{P.~D.~Barnes~Jr.}
%\affiliation{\LLL}

\author{G.~Barr}
\affiliation{\Oxford}

%\author{W.~L.~Barrett}
%\affiliation{\Washington}

%\author{E.~Beall}
%\altaffiliation[Now at\ ]{\Cleveland .}
%\affiliation{\ANL}
%\affiliation{\Minnesota}

%\author{B.~R.~Becker}
%\affiliation{\Minnesota}

%\author{A.~Belias}
%\affiliation{\RAL}

%\author{R.~H.~Bernstein}
%\affiliation{\FNAL}

%\author{M.~Betancourt}
%\affiliation{\Minnesota}

%\author{D.~Bhattacharya}
%\affiliation{\Pittsburgh}

%\author{M.~Bhattarai}
%\affiliation{\Texas}
%\affiliation{\Duluth}

%\author{M.~Bishai}
%\affiliation{\BNL}

\author{A.~Blake}
\affiliation{\Cambridge}
\affiliation{\Lancaster}

%\author{B.~Bock}
%\affiliation{\Duluth}

%\author{G.~J.~Bock}
%\affiliation{\FNAL}

%\author{D.~J.~Boehnlein}
%\affiliation{\FNAL}

%\author{D.~Bogert}
%\affiliation{\FNAL}

%\author{P.~M.~Border}
%\affiliation{\Minnesota}

%\author{C.~Bower}
%\affiliation{\Indiana}

%\author{E.~Buckley-Geer}
%\affiliation{\FNAL}

\author{S.~V.~Cao}
\affiliation{\Texas}

\author{T.~J.~Carroll}
\altaffiliation[Now at\ ]{\Wisconsin}
\affiliation{\Texas}

\author{C.~M.~Castromonte}
\affiliation{\UFG}

%\author{S.~Cavanaugh}
%\affiliation{\Harvard}

%\author{J.~D.~Chapman}
%\affiliation{\Cambridge}

\author{R.~Chen}
\affiliation{\Manchester}

%\author{D.~Cherdack}
%\affiliation{\Tufts}

\author{S.~Childress}
\affiliation{\FNAL}

%\author{B.~C.~Choudhary}
%\altaffiliation[Now at\ ]{\Delhi .}
%\affiliation{\FNAL}
%\affiliation{\Caltech}

\author{J.~A.~B.~Coelho}
\affiliation{\Tufts}

%\author{J.~H.~Cobb}
%\affiliation{\Oxford}

%\author{S.~J.~Coleman}
%\affiliation{\WandM}

%\author{L.~Corwin}
%\altaffiliation[Now at\ ]{\SDakota .}
%\affiliation{\Indiana}

%\author{J.~P.~Cravens}
%\affiliation{\Texas}

%\author{D.~Cronin-Hennessy}
%\affiliation{\Minnesota}

%\author{A.~J.~Culling}
%\affiliation{\Cambridge}

%\author{I.~Z.~Danko}
%\affiliation{\Pittsburgh}

%\author{J.~K.~de~Jong}
%\affiliation{\Oxford}
%\affiliation{\IIT}

\author{S.~De~Rijck}
\affiliation{\Texas}

%\author{A.~V.~Devan}
%\affiliation{\WandM}

%\author{N.~E.~Devenish}
%\affiliation{\Sussex}

%\author{M.~Dierckxsens}
%\affiliation{\BNL}

%\author{M.~V.~Diwan}
%\affiliation{\BNL}

%\author{M.~Dorman}
%\affiliation{\UCL}
%\affiliation{\RAL}

%\author{D.~Drakoulakos}
%\affiliation{\Athens}

%\author{T.~Durkin}
%\affiliation{\RAL}

%\author{S.~A.~Dytman}
%\affiliation{\Pittsburgh}

%\author{A.~R.~Erwin}
%\affiliation{\Wisconsin}

%\author{C.~O.~Escobar}
%\affiliation{\UNICAMP}

\author{J.~J.~Evans}
\affiliation{\Manchester}

%\affiliation{\UCL}
%\affiliation{\Oxford}

%\author{E.~Falk}
%\affiliation{\Sussex}

\author{G.~J.~Feldman}
\affiliation{\Harvard}

%\author{T.~H.~Fields}
%\affiliation{\ANL}

\author{W.~Flanagan}
\affiliation{\Dallas}

%\author{R.~Ford}
%\affiliation{\FNAL}

%\author{M.~V.~Frohne}
%\altaffiliation[Now at\ ]{\HolyCross .}
%\altaffiliation{\deceased}
%\affiliation{\HolyCross}
%\affiliation{\Benedictine}

\author{M.~Gabrielyan}
\affiliation{\Minnesota}

%\author{H.~R.~Gallagher}
%\affiliation{\Tufts}
%\affiliation{\Oxford}
%\affiliation{\ANL}
%\affiliation{\Minnesota}

\author{S.~Germani}
\affiliation{\UCL}

%\author{A.~Godley}
%\affiliation{\Carolina}

%\author{J.~Gogos}
%\affiliation{\Minnesota}

\author{R.~A.~Gomes}
\affiliation{\UFG}

%\author{M.~C.~Goodman}
%\affiliation{\ANL}

\author{P.~Gouffon}
\affiliation{\USP}

\author{N.~Graf}
%\affiliation{\IIT}
\affiliation{\Pittsburgh}

%\author{R.~Gran}
%\affiliation{\Duluth}

%\author{N.~Grant}
%\affiliation{\RAL}

%\author{E.~W.~Grashorn}
%\altaffiliation[Now at\ ]{\Ohio .}
%\affiliation{\Minnesota}
%\affiliation{\Duluth}

%\author{N.~Grossman}
%\affiliation{\FNAL}

\author{K.~Grzelak}
\affiliation{\Warsaw}
%\affiliation{\Oxford}

\author{A.~Habig}
\affiliation{\Duluth}

\author{S.~R.~Hahn}
\affiliation{\FNAL}

%\author{D.~Harris}
%\affiliation{\FNAL}

%\author{P.~G.~Harris}
%\affiliation{\Sussex}

\author{J.~Hartnell}
\affiliation{\Sussex}
%\affiliation{\RAL}
%\affiliation{\Oxford}

%\author{E.~P.~Hartouni}
%\affiliation{\LLL}

\author{R.~Hatcher}
\affiliation{\FNAL}

%\author{K.~Heller}
%\affiliation{\Minnesota}

%\author{A.~Himmel}
%\affiliation{\Caltech}

\author{A.~Holin}
\affiliation{\UCL}

%\author{C.~Howcroft}
%\affiliation{\Caltech}
%\affiliation{\Cambridge}

%\author{X.~Huang}
%\affiliation{\ANL}

\author{J.~Huang}
\affiliation{\Texas}

%\author{L.~Hsu}
%\affiliation{\FNAL}

%\author{J.~Hylen}
%\affiliation{\FNAL}

%\author{J.~Ilic}
%\affiliation{\RAL}

%\author{D.~Indurthy}
%\affiliation{\Texas}

%\author{G.~M.~Irwin}
%\affiliation{\Stanford}

%\author{M.~Ishitsuka}
%\affiliation{\Indiana}

%\author{Z.~Isvan}
%\affiliation{\BNL}
%\affiliation{\Pittsburgh}

%\author{D.~E.~Jaffe}
%\affiliation{\BNL}

%\author{C.~James}
%\affiliation{\FNAL}

%\author{D.~Jensen}
%\affiliation{\FNAL}

%\author{T.~Kafka}
%\affiliation{\Tufts}

%\author{H.~J.~Kang}
%\affiliation{\Stanford}

%\author{S.~M.~S.~Kasahara}
%\affiliation{\Minnesota}

%\author{J.~J.~Kim}
%\affiliation{\Carolina}

%\author{M.~S.~Kim}
%\affiliation{\Pittsburgh}

\author{L.~W.~Koerner}
\affiliation{\Houston}
%\affiliation{\BNL}

%\author{G.~Koizumi}
%\affiliation{\FNAL}

%\author{S.~Kopp}
%\affiliation{\Texas}

\author{M.~Kordosky}
\affiliation{\WandM}
%\affiliation{\UCL}
%\affiliation{\Texas}

%\author{K.~Korman}
%\affiliation{\Duluth}

%\author{D.~J.~Koskinen}
%\altaffiliation[Now at\ ]{\PennState .}
%\affiliation{\UCL}
%\affiliation{\Duluth}

%\author{S.~K.~Kotelnikov}
%\affiliation{\Lebedev}

%\author{Z.~Krahn}
%\affiliation{\Minnesota}

\author{A.~Kreymer}
\affiliation{\FNAL}

%\author{S.~Kumaratunga}
%\affiliation{\Minnesota}

\author{K.~Lang}
\affiliation{\Texas}

%\author{R.~Lee}
%\altaffiliation[Now at\ ]{\MIT .}
%\affiliation{\Harvard}

%\author{G.~Lefeuvre}
%\affiliation{\Sussex}

%\author{J.~Ling}
%\affiliation{\BNL}
%\affiliation{\Carolina}

%\author{P.~J.~Litchfield}
%\affiliation{\Minnesota}
%\affiliation{\RAL}

%\author{R.~P.~Litchfield}
%\affiliation{\Oxford}

%\author{L.~Loiacono}
%\affiliation{\Texas}

\author{P.~Lucas}
\affiliation{\FNAL}

\author{W.~A.~Mann}
\affiliation{\Tufts}

%\author{A.~Marchionni}
%\affiliation{\FNAL}

\author{M.~L.~Marshak}
\affiliation{\Minnesota}

%\author{J.~S.~Marshall}
%\affiliation{\Cambridge}

%\author{M.~Mathis}
%\affiliation{\WandM}

\author{N.~Mayer}
\affiliation{\Tufts}
%\affiliation{\Indiana}
%\affiliation{\Duluth}

%\author{C.~McGivern}
%\affiliation{\Pittsburgh}

%\author{A.~M.~McGowan}
%\altaffiliation[Now at\ ]{\Rochester .}
%\affiliation{\ANL}
%\affiliation{\Minnesota}

%\author{M.~M.~Medeiros}
%\affiliation{\UFG}

\author{R.~Mehdiyev}
\affiliation{\Texas}

\author{J.~R.~Meier}
\affiliation{\Minnesota}

%\author{G.~I.~Merzon}
%\affiliation{\Lebedev}

%\author{M.~D.~Messier}
%\affiliation{\Indiana}
%\affiliation{\Harvard}

%\author{C.~J.~Metelko}
%\affiliation{\RAL}

%author{D.~G.~Michael}
%\altaffiliation{\deceased}
%\affiliation{\Caltech}

%\author{R.~H.~Milburn}
%\affiliation{\Tufts}

%\author{J.~L.~Miller}
%\altaffiliation{\deceased}
%\affiliation{\JMU}
%\affiliation{\Indiana}

\author{W.~H.~Miller}
\affiliation{\Minnesota}

\author{G. Mills}
\altaffiliation{\deceased}
\affiliation{\LANL}

%\author{S.~R.~Mishra}
%\affiliation{\Carolina}
%\affiliation{\Harvard}

%\author{A.~Mislivec}
%\affiliation{\Duluth}

%\author{J.~Mitchell}
%\affiliation{\Cambridge}

%\author{S.~Moed~Sher}
%\affiliation{\FNAL}

%\author{C.~D.~Moore}
%\affiliation{\FNAL}

%\author{J.~Morf\'{i}n}
%\affiliation{\FNAL}

%\author{L.~Mualem}
%\affiliation{\Caltech}
%\affiliation{\Minnesota}

%\author{S.~Mufson}
%\affiliation{\Indiana}

%\author{S.~Murgia}
%\affiliation{\Stanford}

%\author{J.~Musser}
%\affiliation{\Indiana}

\author{D.~Naples}
\affiliation{\Pittsburgh}

\author{J.~K.~Nelson}
\affiliation{\WandM}
%\affiliation{\FNAL}
%\affiliation{\Minnesota}

%\author{H.~B.~Newman}
%\affiliation{\Caltech}

\author{R.~J.~Nichol}
\affiliation{\UCL}

%\author{T.~C.~Nicholls}
%\affiliation{\RAL}

%\author{J.~A.~Nowak}
%\altaffiliation[Now at\ ]{\Lancaster .}
%\affiliation{\Minnesota}

%\author{J.~P.~Ochoa-Ricoux}
%\altaffiliation[Now at\ ]{\Berkeley .}
%\affiliation{\Caltech}

\author{J.~O'Connor}
\affiliation{\UCL}

%\author{W.~P.~Oliver}
%\affiliation{\Tufts}

%\author{M.~Orchanian}
%\affiliation{\Caltech}

%\author{T.~Osiecki}
%\affiliation{\Texas}

%\author{R.~Ospanov}
%\altaffiliation[Now at\ ]{\PennU .}
%\affiliation{\Texas}

%\author{S.~Osprey}
%\affiliation{\Oxford}

\author{R.~B.~Pahlka}
\affiliation{\FNAL}

%\author{J.~Paley}
%\affiliation{\ANL}
%\affiliation{\Indiana}

%\author{V.~Paolone}
%\affiliation{\Pittsburgh}

%\author{A.~Para}
%\affiliation{\FNAL}

%\author{R.~B.~Patterson}
%\affiliation{\Caltech}

%\author{T.~Patzak}
%\affiliation{\CdF}
%\affiliation{\Tufts}

\author{\v{Z}.~Pavlovi\'{c}}
\affiliation{\FNAL}
%\affiliation{\Texas}

%\author{\v{Z}.~Pavlovi\'{c}}
%\altaffiliation[Now at\ ]{\FNAL .}
%\affiliation{\LANL}
%\affiliation{\Texas}

\author{G.~Pawloski}
\affiliation{\Minnesota}
%\affiliation{\Stanford}

%\author{G.~F.~Pearce}
%\affiliation{\RAL}

%\author{C.~W.~Peck}
%\affiliation{\Caltech}

\author{A.~Perch}
\affiliation{\UCL}

%\author{E.~A.~Peterson}
%\affiliation{\Minnesota}

%\author{D.~A.~Petyt}
%\affiliation{\Minnesota}
%\affiliation{\RAL}
%\affiliation{\Oxford}

\author{M.~M.~Pf\"{u}tzner}  % fixed 02/12/16
\affiliation{\UCL}

\author{D.~D.~Phan}
\affiliation{\Texas}

%\author{S.~Phan-Budd}
%\affiliation{\ANL}

%\author{H.~Ping}
%\affiliation{\Wisconsin}

%\author{R.~Pittam}
%\affiliation{\Oxford}

\author{R.~K.~Plunkett}
\affiliation{\FNAL}

\author{N.~Poonthottathil}
\affiliation{\FNAL}

\author{X.~Qiu}
\affiliation{\Stanford}

\author{A.~Radovic}
\affiliation{\WandM}

%\author{D.~Rahman}
%\affiliation{\Minnesota}

%\author{A.~Rahaman}
%\affiliation{\Carolina}

%\author{R.~A.~Rameika}
%\affiliation{\FNAL}

%\author{J.~Ratchford}
%\affiliation{\Texas}

%\author{T.~M.~Raufer}
%\affiliation{\RAL}
%\affiliation{\Oxford}

%\author{B.~Rebel}
%\affiliation{\FNAL}
%\affiliation{\Indiana}

%\author{J.~Reichenbacher}
%\altaffiliation[Now at\ ]{\Alabama .}
%\affiliation{\ANL}

%\author{D.~E.~Reyna}
%\affiliation{\ANL}

%\author{P.~A.~Rodrigues}
%\affiliation{\Oxford}

%\author{C.~Rosenfeld}
%\affiliation{\Carolina}

%\author{H.~A.~Rubin}
%\affiliation{\IIT}

%\author{K.~Ruddick}
%\affiliation{\Minnesota}

%\author{V.~A.~Ryabov}
%\affiliation{\Lebedev}

%\author{R.~Saakyan}
%\affiliation{\UCL}

\author{P.~Sail}
\affiliation{\Texas}

\author{M.~C.~Sanchez}
\affiliation{\Iowa}
%\affiliation{\ANL}
%\affiliation{\Harvard}
%\affiliation{\Tufts}

%\author{N.~Saoulidou}
%\affiliation{\FNAL}
%\affiliation{\Athens}

\author{J.~Schneps}
\altaffiliation{\deceased}
\affiliation{\Tufts}

\author{A.~Schreckenberger}
\affiliation{\Texas}
%\affiliation{\Minnesota}

%\author{P.~Schreiner}
%\affiliation{\ANL}

%\author{V.~K.~Semenov}
%\affiliation{\IHEP}

%\author{S.-M.~Seun}
%\affiliation{\Harvard}

%\author{P.~Shanahan}
%\affiliation{\FNAL}

\author{R.~Sharma}
\affiliation{\FNAL}

%\author{W.~Smart}
%\affiliation{\FNAL}

%\author{V.~Smirnitsky}
%\affiliation{\ITEP}

%\author{C.~Smith}
%\affiliation{\UCL}
%\affiliation{\Sussex}
%\affiliation{\Caltech}

\author{A.~Sousa}
\affiliation{\Cincinnati}
%\affiliation{\Harvard}
%\affiliation{\Oxford}
%\affiliation{\Tufts}

%\author{B.~Speakman}
%\affiliation{\Minnesota}

%\author{P.~Stamoulis}
%\affiliation{\Athens}

%\author{M.~Strait}
%\affiliation{\Minnesota}

%\author{P.~Symes}
%\affiliation{\Sussex}

\author{N.~Tagg}
\affiliation{\Otterbein}
%\affiliation{\Tufts}
%\affiliation{\Oxford}

%\author{R.~L.~Talaga}
%\affiliation{\ANL}

%\author{E.~Tetteh-Lartey}
%\affiliation{\TexasAM}

%\author{M.~A.~Tavera}
%\affiliation{\Sussex}

\author{J.~Thomas}
\affiliation{\UCL}
%\affiliation{\Oxford}
%\affiliation{\FNAL}

%\author{J.~Thompson}
%\altaffiliation{\deceased}
%\affiliation{\Pittsburgh}

\author{M.~A.~Thomson}
\affiliation{\Cambridge}

%\author{J.~L.~Thron}
%\altaffiliation[Now at\ ]{\LANL .}
%\affiliation{\ANL}

%\author{X.~Tian}
%\affiliation{\Carolina}

\author{A.~Timmons}
\affiliation{\Manchester}

%\author{G.~Tinti}
%\affiliation{\Oxford}

\author{J.~Todd}
\affiliation{\Cincinnati}

\author{S.~C.~Tognini}
\affiliation{\UFG}

\author{R.~Toner}
\affiliation{\Harvard}
%\affiliation{\Cambridge}

\author{D.~Torretta}
\affiliation{\FNAL}

%\author{I.~Trostin}
%\affiliation{\ITEP}

%\author{V.~A.~Tsarev}
%\affiliation{\Lebedev}

%\author{G.~Tzanakos}
%\altaffiliation{\deceased}
%\affiliation{\Athens}

%\author{J.~Urheim}
%\affiliation{\Indiana}
%\affiliation{\Minnesota}

\author{P.~Vahle}
\affiliation{\WandM}
%\affiliation{\UCL}
%\affiliation{\Texas}

%\author{V.~Verebryusov}
%\affiliation{\ITEP}

%\author{B.~Viren}
%\affiliation{\BNL}

%\author{J.~J.~Walding}
%\affiliation{\WandM}

%\author{C.~P.~Ward}
%\affiliation{\Cambridge}

%\author{D.~R.~Ward}
%\affiliation{\Cambridge}

%\author{M.~Watabe}
%\affiliation{\TexasAM}

\author{A.~Weber}
\affiliation{\Oxford}
\affiliation{\RAL}

%\author{R.~C.~Webb}
%\affiliation{\TexasAM}

%\author{A.~Wehmann}
%\affiliation{\FNAL}

%\author{N.~West}
%\affiliation{\Oxford}

%\author{C.~White}
%\affiliation{\IIT}

\author{L.~H.~Whitehead}
\affiliation{\UCL}

\author{S.~G.~Wojcicki}
\affiliation{\Stanford}

%\author{D.~M.~Wright}
%\affiliation{\LLL}

%\author{T.~Yang}
%\affiliation{\Stanford}

%\author{H.~Zheng}
%\affiliation{\Caltech}

%\author{M.~Zois}
%\affiliation{\Athens}

%\author{K.~Zhang}
%\affiliation{\BNL}

%\author{R.~Zwaska}
%\affiliation{\FNAL}

\collaboration{The MINOS+ Collaboration}
\noaffiliation

\date{\today}

\begin{abstract}
We report the final measurement of the neutrino oscillation parameters $\Delta m^{2}_{32}$ and $\sin^{2}\theta_{23}$ using all data from the
MINOS and MINOS+ experiments. These data were collected using a total exposure of $23.76 \times 10^{20}$\,protons on target producing \numu and \numubar beams and \unit[60.75]{kt$\cdot$yr} exposure to atmospheric neutrinos. The measurement of the disappearance of \numu  and the appearance of \nue events between the Near and Far  detectors yields $|\Delta m^2_{32}|=2.40^{+0.08}_{-0.09}~(2.45^{+0.07}_{-0.08}) \times 10^{-3}$\,eV$^2$ and $\sin^2\theta_{23} = 0.43^{+0.20}_{-0.04} ~(0.42^{+0.07}_{-0.03})$ at 68\% C.L. for Normal (Inverted) Hierarchy.

\end{abstract}


%\keywords{Suggested keywords}%Use showkeys class option if keyword display desired

\maketitle
Since the discovery of $\nu_{\mu}$ flavor disappearance oscillations in atmospheric neutrinos by the SuperKamiokande experiment in 1998~\cite{Fukuda:1998mi}, determinations of neutrino mass-squared differences and mixing angles have steadily improved \cite{KamLANDlatest,SNOlatest,ref:MINOSCombinedPRL,IceCubelatest,T2Klatest,Abe:2019vii,SKlatest,DayaBaylatest,RENOlatest,NOvAlatest,ref:MINOSNuNuBarPRL2013}, but the need for more precision remains. This is especially the case for the atmospheric $\theta_{23}$ mixing angle. A value corresponding to maximal mixing, $\theta_{23}$ = 45\,$^\circ$, may be a harbinger for an underlying symmetry. On the other hand, if the mixing is non-maximal, determination of its octant is important for $\nu_{e}$-flavor appearance measurements and the inference of the CP violating angle $\delta_{CP}$\cite{T2Klatest}.

This Letter reports new measurements of $\Delta m_{32}^{2}$ and $\sin^{2}(\theta_{23})$ using the complete set of beam and atmospheric data taken with the MINOS detectors.
%from 2005 to 2016. 
Two distinct beam energy configurations of the NuMI neutrino beam at Fermi National Accelerator Laboratory (FNAL) corresponded to two phases of the MINOS (2005-2012) and MINOS+ (2013-2016) long-baseline, on-axis neutrino oscillation experiments. The MINOS+ dataset significantly increases the statistics of the MINOS measurements~\cite{ref:MINOSCombinedPRL} in the energy region above the oscillation maximum in the standard model of oscillations. Over this region extending from 1.5 \unit{GeV} to above 10 \unit{GeV} in E$_{\nu}$, MINOS and MINOS+ monitor the increase of $\nu_{\mu}$-flavor survival probability. This provides additional sensitivity, not available to narrow-band beam experiments, for measuring the extent to which $\theta_{23}$ deviates from non-maximal mixing. Monitoring this revival rate supplements the measurement of the depth of the oscillation maximum which occurs within a small span of E$_{\nu}$. Effects from nonstandard neutrino interactions~\cite{NSI}, neutrino decay~\cite{Barger:1999bg,ref:MINOSPRL2011}, decoherence~\cite{decoherence}, or the existence of sterile neutrinos~\cite{ref:MINOSSterile,LED}, could manifest themselves over the large energy range. Consequently, the measurements reported here will also allow a stringent test for such phenomena as well as future hypotheses, that lie outside the purview of conventional three-flavor neutrino oscillations.

The MINOS and MINOS+ long-baseline, on-axis neutrino oscillation experiments recorded two distinct phases of exposure to the NuMI neutrino beam \cite{numibeam} at Fermilab utilizing the MINOS Near and Far detectors\cite{minos}.  Both detectors were functionally equivalent magnetized steel-scintillator, tracking, sampling calorimeters. The Near Detector (ND) was 1.04\,km from the target, 103\,m underground, and had a mass of 980\,t. The Far Detector (FD) was 735 km from the target, 705\,m underground, and had a mass of 5.4\,kt. The detectors had average toroidal magnetic fields of 1.4\,T, to enable the separation of $\nu_{\mu}$ from $\overline{\nu}_{\mu}$. The FD was also used to study atmospheric neutrinos~\cite{ref:minosatmosphericprd} making use of the scintillator veto shield to improve cosmic muon background rejection.

During MINOS-phase data taking, the NuMI beam operated primarily in a low-energy beam configuration, producing muon neutrinos or antineutrinos, depending on the polarity of the pulsed magnetic horns, with a peak energy around \unit[3]{GeV}. The MINOS low-energy beam exposure was $10.56 \times 10^{20}$\,protons on target (POT) in $\numu$-mode and $3.36 \times 10^{20}$\,POT in $\numubar$-mode. The MINOS $\numu$-mode sample included an additional \unit[$0.15 \times 10^{20}$]{POT} exposure in a high-energy $\numu$-mode with a peak \numu energy of \unit[9]{GeV}. The analysis of this MINOS-phase data has been described previously, presenting a $\unit[37.88]{kt\cdot yr}$ sample of atmospheric neutrinos, measurements of both \numu and \numubar disappearance, and \nue and \nuebar appearance~\cite{ref:minosatmosphericprd,ref:MINOSNuNuBarPRL2013,ref:MINOSNuEPRL2013,ref:MINOSCombinedPRL}.

This analysis uses the complete MINOS data set described above and, in addition, includes $\unit[22.87]{kt\cdot yr}$ of atmospheric-neutrino data from 2011--2016, along with the complete three years of MINOS+ $\numu$-mode beam data corresponding to an exposure of $9.69 \times 10^{20}$\,POT. In the MINOS+ phase, the NuMI beam operated in the medium-energy configuration, producing a \numu beam peaking at near \unit[7]{GeV}. The MINOS+ \numu  charged current (CC) interactions in the ND were composed of 96.9\% \numu, 1.9\% \numubar, and 1.2\% ($\nue+\nuebar$). In comparison, the MINOS low-energy \numu-mode ND data were composed of 92.9\% \numu, 5.8\% \numubar, and 1.3\% ($\nue+\nuebar$) CC interactions~\cite{minosbeamprd}.  There is no \nue appearance included in the MINOS+ phase analysis since the higher energy exposure increases the neutral-current (NC) backgrounds to the low energy \nue appearance signal.

The Monte Carlo (MC) modeling was unchanged compared to the most recent previous publications. The accelerator beam neutrino flux was simulated using the FLUGG package~\cite{ref:FLUGG} and the atmospheric neutrino flux using the Bartol calculations~\cite{ref:Bartol}.
Beam and atmospheric neutrino interactions in the detector are simulated using NEUGEN3~\cite{ref:NEUGEN3}, and interactions of atmospheric neutrinos in the surrounding rock are propagated into the detector using NUANCE~\cite{ref:NUANCE}.
 The detector response to final-state particles is simulated, for both beam and atmospheric neutrino interactions, using a combination of GEANT3~\cite{ref:GEANT3} and GCALOR~\cite{ref:GCALOR}. For MINOS+, the beam-neutrino reconstruction algorithms were tuned to account for the higher occupancy in the ND arising from the increased beam-neutrino flux. The data observed in the ND are used to tune the MINOS and MINOS+ flux simulations. The MINOS flux tuning procedure, described previously~\cite{minosbeamprd}, was improved upon to separate the effect of the rate of secondary hadron production in the target from that of the charged particle focussing by the horns. This procedure combined data from special ND data taken with horn currents between 0 and 200~kA.

During the MINOS+ running period, it was observed that the neutrino energy peak position was shifted in the ND from that predicted by the MC by about 400\,MeV. Furthermore, during the final running period, the upstream support of the first magnetic horn moved downward \unit[4]{mm} over a period of a few months. Checks on these effects, however, showed that the oscillation parameter measurement is robust against these MC/data differences once the ND data are used to correct the neutrino flux to less than $0.05\sigma$ in both oscillation parameters.

\begin{figure}
\begin{center}
\includegraphics[width=\columnwidth]{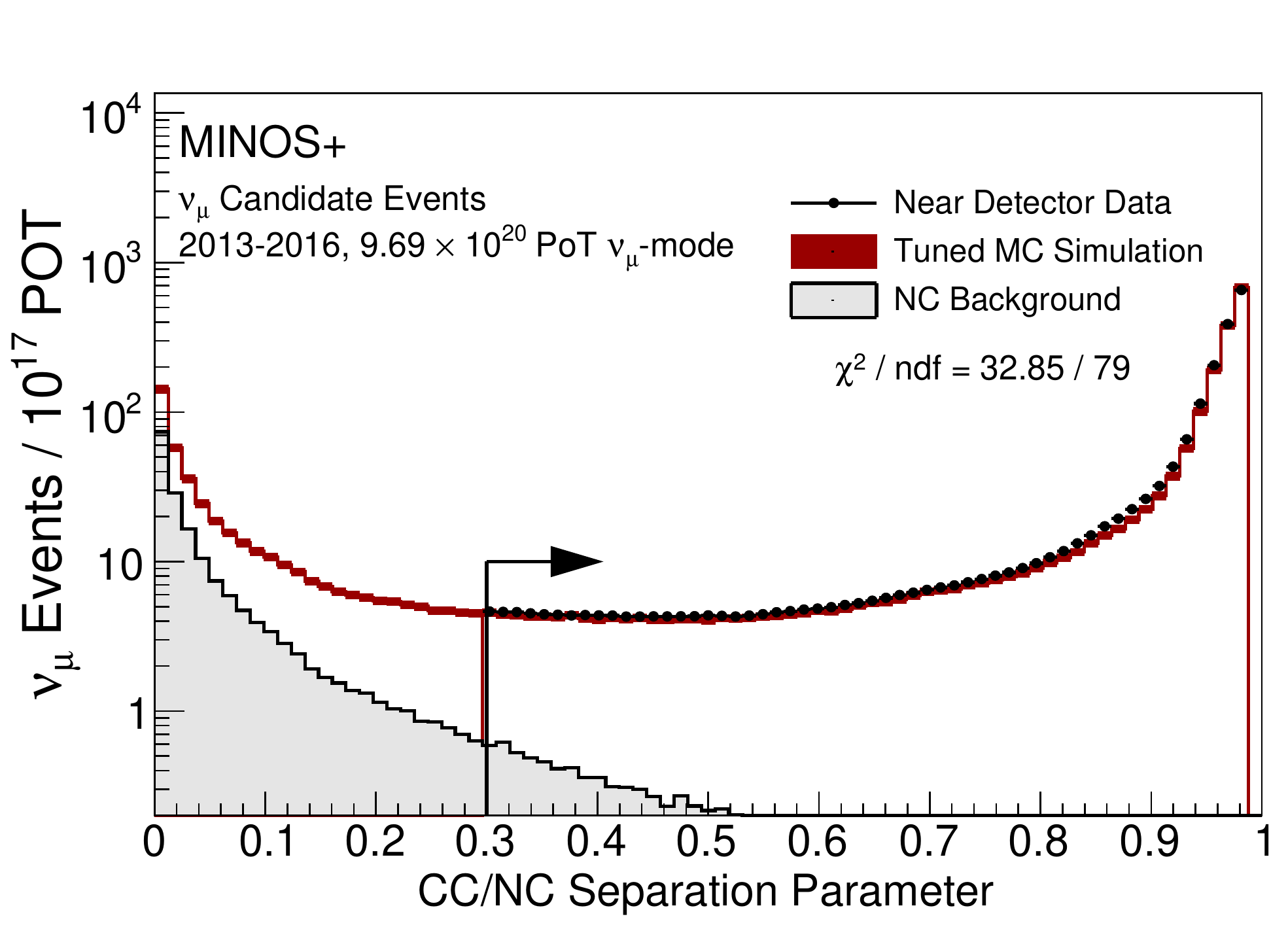} 
\end{center}
\caption{The $k$NN discriminant used in MINOS+ data to separate CC \numu and \numubar interactions from the background of NC interactions. The discriminant is shown for data and MC in the ND, where the MC has been corrected using the beam tuning described in the text. The red band is the systematic uncertainty on the MC prediction yielding a $\chi^{2}/d.o.f$ of (32.85/79) for the data/MC comparison. Interactions in both detectors with the discriminant value above 0.3 are selected by the analysis.}
\label{fig:kNN}
\end{figure}

 This analysis uses \numu and \numubar events, which result in a $\mu^{-}$ or $\mu^{+}$ in the final state. Signal events have a characteristic muon track with a hadron shower near the interaction point. The major source of background is from NC events that produce hadron showers  with short tracks. A multivariate $k$-Nearest Neighbor ($k$NN) algorithm \cite{kNN:6} was used by MINOS to select CC $\numu$ and $\numubar$ interactions based on event topology and the characteristic muon track energy deposition. For MINOS+, the algorithm was trained using representative MINOS+ CC and NC events from the MC~\cite{Junting}; the distribution of the $k$NN discriminant in MINOS+ ND MC is shown in Fig.~\ref{fig:kNN}. Events with a value of the $k$NN discriminant below 0.3 were removed. The selected CC \numu and \numubar sample has a purity of 99.1\% in the ND and 99.3\% in the FD, with the impurities due to NC interactions.

The visible energy of the selected \numu and \numubar events was reconstructed from the sum of the muon track and the hadronic shower energy. The muon energy was measured from the range in the detector for tracks that were fully contained in the detector, or from the curvature in the magnetic field for tracks that exited the detector. The shower energy was estimated using another $k$NN algorithm that compares the topology of the event to a library of MC events and uses the closest-matching MC events to estimate the true energy of the hadron shower~\cite{Junting,Backhouse}.

The beam data consist of \numu and \numubar events with reconstructed interaction vertices within the detector's fiducial volume. The MINOS \numu dataset includes a sample of non-fiducial muons from neutrinos that interacted outside the detector's fiducial volume or in the rock surrounding the detector, identified by muons entering the front or sides of the detector in time with the beam. Since this sample has poorly reconstructed interaction energy, its impact on the oscillation measurement is limited and thus did not warrant selecting a similar sample from
the MINOS+ data \cite{ref:MattStrait}.

\begin{figure}
\begin{center}
{\includegraphics[width=\columnwidth]{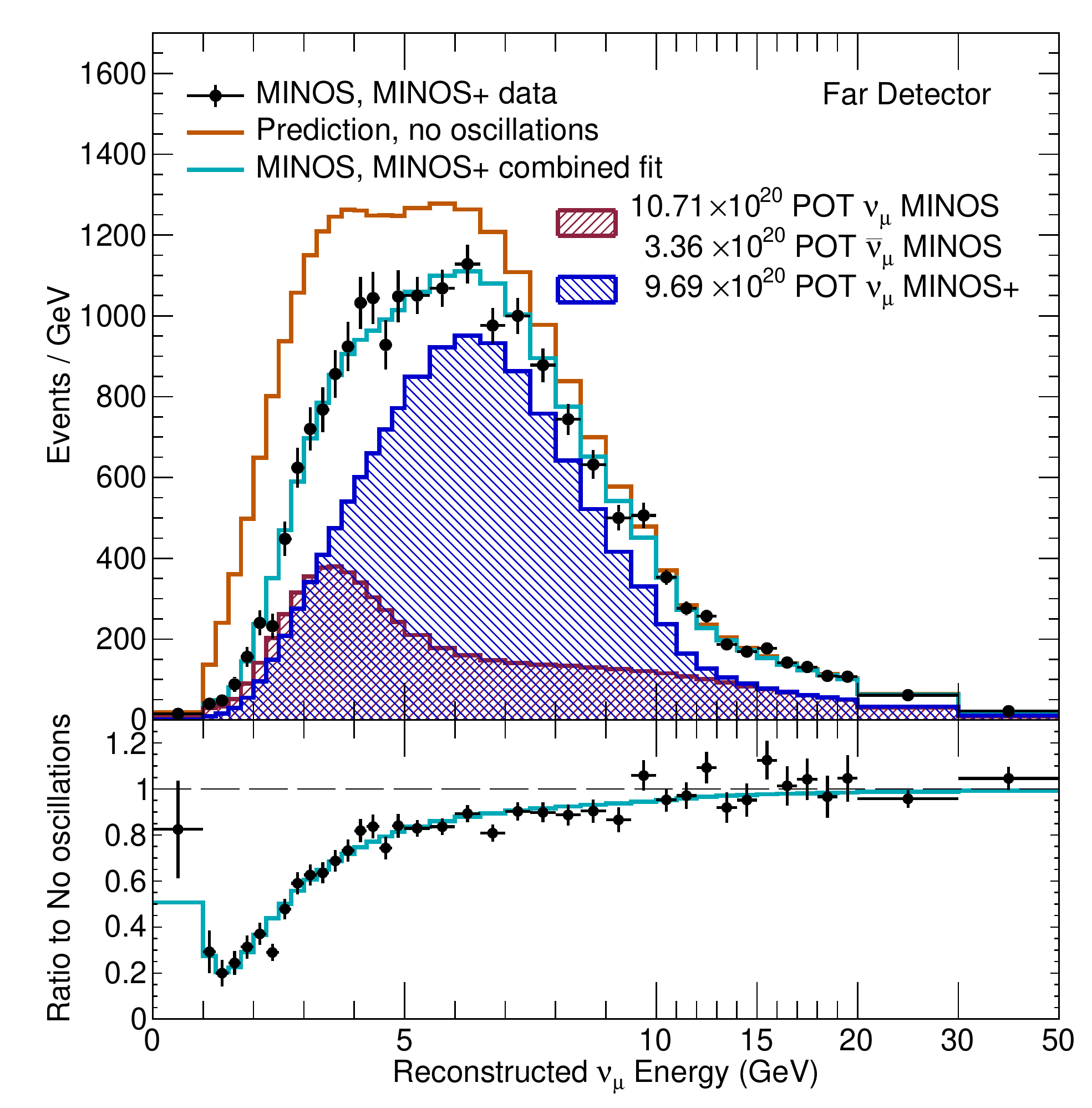}\label{fig:FDSpectra}}%
\end{center}%
\caption{Top: The reconstructed energy spectra for MINOS and MINOS+ events selected within the Far Detector fiducial volume for data (black points) and the best fit MC predictions for MINOS (red hatched histogram), MINOS+ (blue hatched histogram) and the sum (cyan line). The prediction at the FD with no oscillations is shown as the orange line. Bottom: The ratios of the data and the oscillated prediction to the no oscillation prediction for MINOS and MINOS+ combined.}
\label{fig:FDSpectraAndRatio}
\end{figure}

The MINOS+ reconstructed CC \numu and \numubar energy spectra in the ND were used to predict the energy spectrum expected in the FD using a beam transfer matrix, as was done for MINOS~\cite{minosbeamprd}. This prediction for MINOS and MINOS+ combined without oscillations is shown in Fig.~\ref{fig:FDSpectraAndRatio} (orange line) compared to the selected data (black points). Also shown is 
 the ratio of observed FD events to the number of predicted events assuming no oscillations as a function of reconstructed neutrino energy. The energy-dependent deficit of \numu and \numubar interactions is clearly observed, indicating the expected three-flavor oscillatory nature of the disappearance. The MINOS+ data provide significant additional statistical power integrated between \unit[4] and \unit[8]{GeV} energy range over the lower energy MINOS data.

\begin{table}[hbt]
\caption{\label{tab:EventCounts} Selected events in the Far Detector for all MINOS and MINOS+ beam and atmospheric neutrino samples.}
\begin{ruledtabular}
\begin{tabular}{l   c cr  rc}
& \multicolumn{3}{c}{Data} & \multicolumn{2}{c}{Predicted}\\
&  MINOS & MINOS+ & Total&  \ No. Osc.  & Best Fit \\
\colrule
	$\numu$ ($\numu$ beam)      &  2579   & 6280  &8859& 10634 & 8851 \\ 
    $\numubar$ ($\numu$ beam)   &  \ 312   & 293    &605& 677 & \ 598 \\
    Nonfid. $\mu^- + \mu^+$             &  2911   &N/A&   2911    & 3256 & 2838 \\
    $\numubar$ ($\numubar$ beam)&  \ 226    &    N/A &226& 320 & \ 225 \\
    Atm. $\numu + \numubar$     &  \ 905    &  473  &1378& 1885 & 1366 \\
    Atm. $\mu^- + \mu^+$        &  \ 466    &  270&736  & 930   & \ 737 \\
    Atm. showers                &  \ 701    &  422&1123  & 1224 & 1130 \\
\end{tabular}
\end{ruledtabular}
\end{table}

Atmospheric neutrinos were separated from the cosmic ray backgrounds into three separate samples~\cite{ref:minosatmosphericprd,chapmanthesis,speakmanthesis,perchthesis}. The first sample of \numu and \numubar CC interactions required a reconstructed interaction vertex in the detector's fiducial volume. The second sample were also contained-vertex, but shower-like events, primarily CC \nue, CC \nuebar, and NC interactions that were used to constrain the atmospheric neutrino flux. The third sample required a reconstructed upward-going muon track and contained nonfiducial events that were initiated by atmospheric \numu and \numubar interactions that occurred in the rock around the detector.  The number of observed and predicted neutrino events for MINOS and MINOS+ are given in Table~\ref{tab:EventCounts}.

The combined fit to the MINOS and MINOS+ \numu disappearance data was carried out independently of the MINOS \nue appearance fit. To determine values of the oscillation parameters from the muon neutrino data, a maximum likelihood fit was performed by varying $\Delta m^2_{32}$, $\sin^2\theta_{23}$, $\sin^2\theta_{13}$, and $\delta_{\rm CP}$ and using the negative log-likelihood function:
$$-\ln\mathcal{L} = \sum_j \mu_j - n_j + n_j \ln(n_j/\mu_j) + 0.5\sum_k \left( \frac{\alpha_k}{\sigma_{\alpha_k}} \right )^2,$$ where $\mu_j$ and $n_j$ are the numbers of expected and observed events in bin $j$ of the reconstructed energy distribution, $\alpha_k$ include the fitted systematic parameters and a constraint on $\sin^2\theta_{13}$ with corresponding uncertainties of $\sigma_{\alpha_k}$. The mixing angle $\theta_{13}$ was constrained to ${\sin^2\theta_{13} = 0.0210 \pm 0.0011}$ \cite{pdg}. The solar parameters were fixed to ${\Delta m^2_{21} = 7.54 \times 10^{-5}\,{\rm eV}^2}$ and ${\sin^2\theta_{12} = 0.307}$ \cite{solarRef} since they have no effect on the oscillation parameter measurement. The likelihood function contained 17 nuisance parameters that accounted for the largest systematic uncertainties as discussed in previous publications \cite{ref:MINOSPRL2011,ref:minosatmosphericprd}.

 Uncertainties on the flux of beam neutrinos were obtained from the fits performed using the ND data to tune the flux simulation. Separate uncertainties were calculated for the MINOS and MINOS+ beam-neutrino data sets. All uncertainties related to the interactions of neutrinos in the detector and neutrino reconstruction were unchanged from MINOS to MINOS+. All uncertainties on the atmospheric-neutrino samples were unchanged between the previously analyzed data and the new data added for this publication.

The effects of the systematic uncertainties on the $\nu_{\mu}$ disappearance measurement were studied with MC samples modified by shifting the uncertainties by one standard deviation. Table~\ref{table:systematics} shows the largest of the systematic uncertainties on the $\Delta m^2_{32}$ measurement. The dominant uncertainties associated with the beam data  are the shower energy uncertainty and the relative normalization between the two detectors.  The shower energy uncertainty, which has the second largest impact on $\Delta m^2_{32}$, averages at about 8\% below 3\,GeV and approaches 6.6\% at higher shower energies. The 1.6\% relative normalization uncertainty accounts for differences in event selection and reconstruction between the ND and FD as well as uncertainties on each detector's fiducial mass and live-time. The uncertainty on the measurement of the muon energy is fully correlated between the beam and atmospheric samples and is 2\% (3\%) when calculated from range and 3\% (5\%) when calculated from curvature\cite{ref:MINOSPRL2011} for the beam (atmospheric) samples. The difference between the samples is attributed to the orientation of the detector planes relative to the incident muons.

The 15\% atmospheric normalization uncertainty for contained-vertex (CV) events comes from uncertainties on the flux and the neutrino cross section\cite{ref:minosatmosphericprd}. This normalization uncertainty has the largest effect on $\Delta m^2_{32}$. The atmospheric nonfiducial events have a normalization uncertainty of 25\% due to larger flux uncertainties of the much higher energy cosmic muons associated with this sample. Atmospheric CV events have an additional 10\% uncertainty on the $\bar{\nu}_\mu / \nu_\mu$ ratio. These three uncertainties have the largest effects on measuring $\sin^2\theta_{23}$ in the combined fit of beam and atmospheric data.

\begin{table}[htb]
\caption{\label{table:systematics}Sources of systematic uncertainties with the largest impact on $\Delta m^2_{32}$ and their effect on fitting the oscillation parameters for one standard deviation variations.}
\begin{ruledtabular}
\begin{tabular}{lcc}
\multirow{2}{*}{Uncertainty} & $\delta\left(\Delta m^2_{32}\right)$ & $\delta\left(\sin^2\theta_{23}\right)$ \\
& $\left(10^{-3}~\rm{eV}^2\right)$ &\\
\colrule
Atm. normalization CV (15\%)& 0.067 & 0.071 \\
Beam shower energy & 0.064 & 0.001\\
Beam relative normalization (1.6\%) & 0.049 & 0.002 \\
$\mu^\pm$ energy (range 2\%, curv. 3\%) & 0.048 & 0.003 \\
Atm. normalization nonfid. (25\%)& 0.032 & 0.053\\
Atm. $\bar{\nu}_\mu / \nu_\mu$ ratio CV (10\%)& 0.012 & 0.012\\
\end{tabular}
\end{ruledtabular}
\end{table}

\begin{figure}
    \centering
    \includegraphics[width=0.49\textwidth]{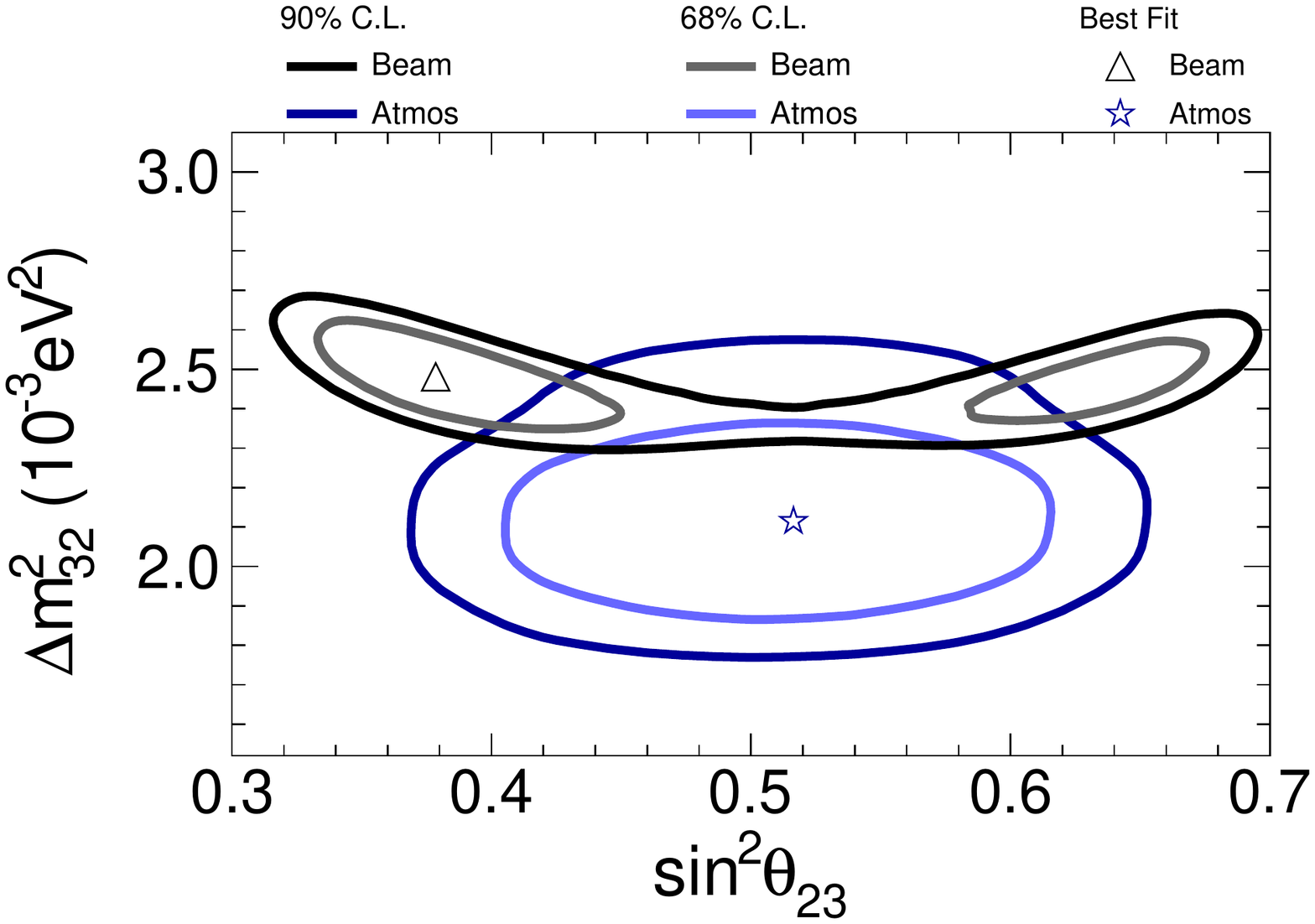}
    \caption{Confidence limits on $\Delta m^2_{32}$ and $\sin^2\theta_{23}$ in the normal hierarchy for MINOS and MINOS+ shown separately for the beam and atmospheric data. Contours are shown at 90\% C.L. and at 68\% C.L. Best fit points are shown for the beam data (triangle) and for the atmospheric fit (star).}
    \label{fig:beam vs atmos}
\end{figure}

The final result sums the likelihood contributions coming separately from the combined $\numu$ disappearance and the MINOS $\nue$ appearance \cite{ref:MINOSNuNuBarPRL2013,ref:MINOSNuEPRL2013} data sets. The \numu disappearance and the \nue appearance analyses use information from the Near Detector in independent ways. The \numu analysis uses ND data to minimize uncertainties in the beam flux and neutrino cross section, while the \nue analysis primarily uses ND data to estimate backgrounds. The dominant uncertainty is statistical; the systematic uncertainties in the two analyses are treated as uncorrelated.

Fig. \ref{fig:FDSpectraAndRatio} shows the MC predictions for the best fit oscillation parameters for MINOS
(hatched red) and MINOS+ (hatched blue). The 
combined MINOS and MINOS+ MC spectrum is also shown (cyan). All MC samples with expected neutrino oscillations include the small contribution of background events from \nutau and \nutaubar appearance. The oscillation parameters best-fit point obtained using only the MINOS+ neutrino beam data falls within the $1\sigma$ contour from the previous MINOS measurement \cite{ref:MINOSCombinedPRL}.

\begin{figure}[hhh]
%\begin{center}
\begin{minipage}{0.5\textwidth}
\vspace{0.6cm}
\begin{minipage}{0.99\textwidth}
\includegraphics[width=\textwidth]{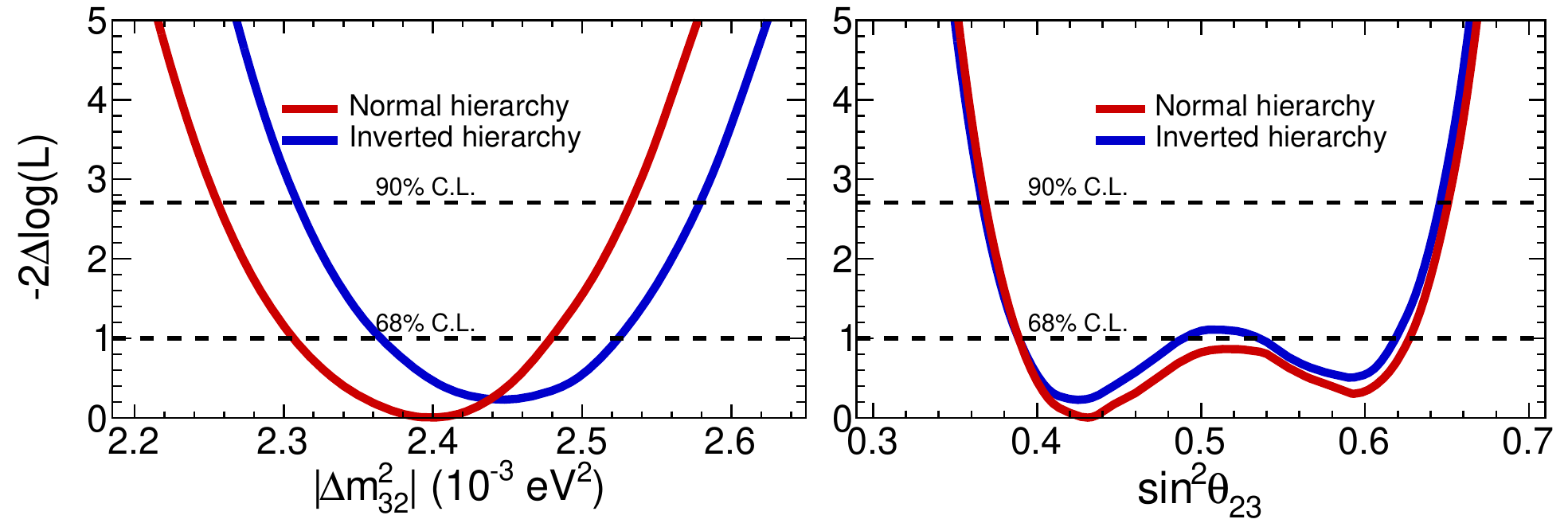}
\end{minipage}
\end{minipage}\hspace{0.5pc}
\begin{center}
\begin{minipage}{0.5\textwidth}
\includegraphics[width=\textwidth]{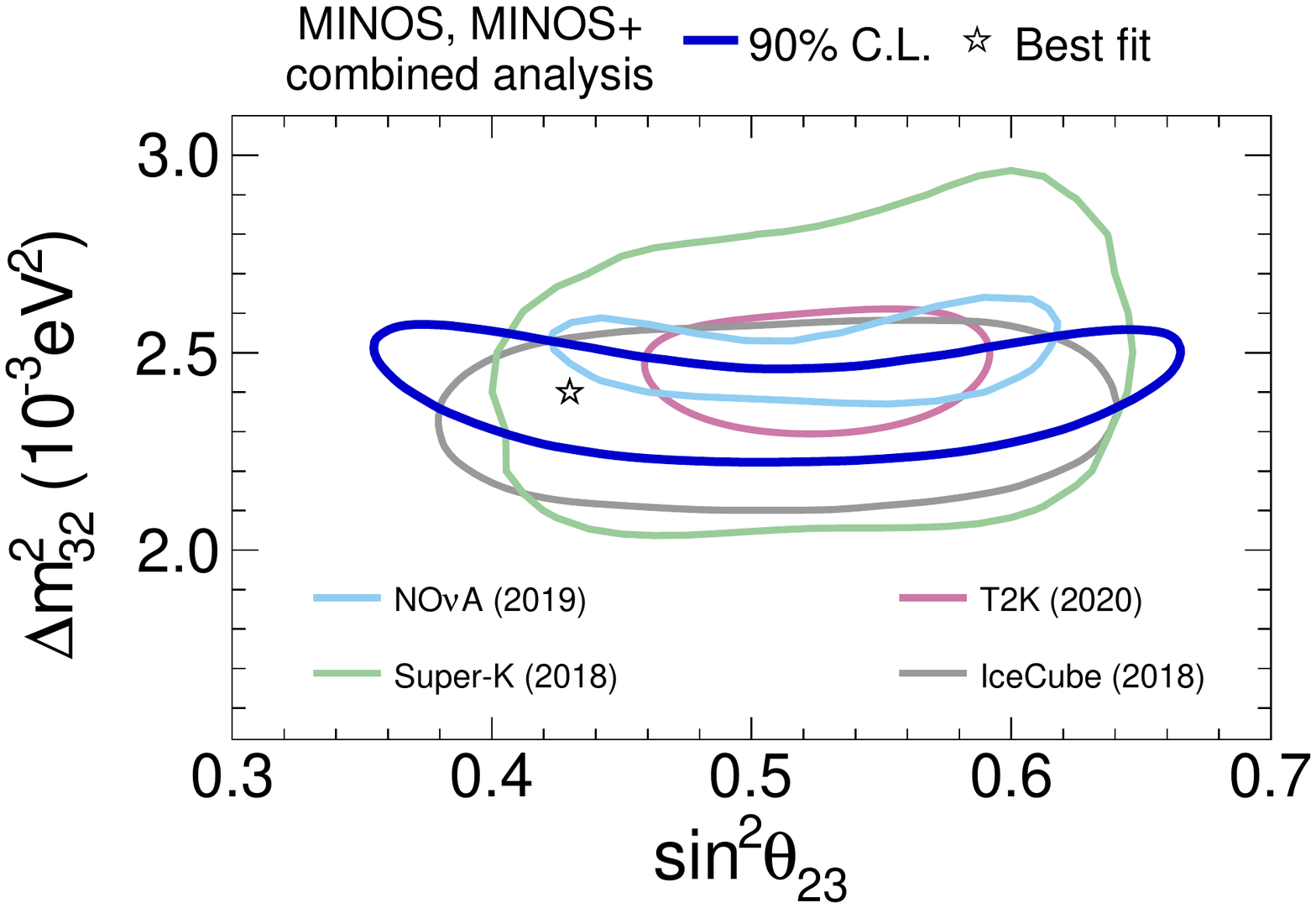}
\end{minipage}
\end{center}
\caption{The top figures show the 1D likelihood profiles as functions of $\Delta m^2_{32}$ and $\sin^2\theta_{23}$ for each hierarchy. The bottom figure displays 90\% confidence limits on $\Delta m^2_{32}$ and $\sin^2\theta_{23}$ for the normal mass hierarchy, comparing MINOS+, IceCube, NO$\nu$A, Super-K, and T2K \cite{NOvAlatest,IceCubelatest,Abe:2019vii,SKlatest}.}
\label{MINOS and MINOS+ Contours}
\end{figure}

\begin{table}[hb]
\caption{\label{bestfits}The best fit values and confidence limits of the $\Delta m^{2}_{32}$ and $\sin^{2}\theta_{23}$ parameters, calculated separately for the normal and inverted hierarchy.
$\Delta m^2_{32}$ is reported in units of $\unit[10^{-3}]{eV^2}$.}
\begin{ruledtabular}
\begin{tabular}{cccc}
Mass Hierarchy & Parameter & Best fit & Confidence limits \\
\colrule
\multirow{2}{*}{Normal} & $|\Delta m^{2}_{32}|$ & 2.40 & 2.31--2.48 (68\% C.L.) \\
& $\sin^{2}\theta_{23}$ & 0.43 & 0.37--0.65 (90\% C.L.)\\ \colrule 
\multirow{2}{*}{Inverted} & $|\Delta m^{2}_{32}|$ & 2.45 & 2.37--2.52 (68\% C.L.) \\
& $\sin^{2}\theta_{23}$ & 0.42 & 0.37--0.65 (90\% C.L.) \\
\end{tabular}
\end{ruledtabular}
\end{table}
The 68\% and 90\% confidence level intervals in $\sin^2\theta_{23}$ and $\Delta m^2_{32}$ parameter space for the normal hierarchy obtained for the beam and atmospheric samples separately are shown in Fig.~\ref{fig:beam vs atmos}.  The confidence level intervals include the best fit points for the beam sample at $\sin^2\theta_{23} = 0.38$, $\Delta m^2_{32} = 2.48 \times 10^{-3}\,\unit{eV^2}$ and for the atmospheric sample at $\sin^2\theta_{23} = 0.52$, ${\Delta m^2_{32} = 2.11 \times 10^{-3}\,\unit{eV^2}}$. Studies of the compatibility of the atmospheric and beam results show a probability of 22\% that they come from the same oscillation parameters.

The oscillation parameters at the best fit point and confidence limits from the overall combined fit for the normal and inverted hierarchy are shown in Table~\ref{bestfits}. Fig.~\ref{MINOS and MINOS+ Contours} shows the confidence limits on $\Delta m^2_{32}$ and $\sin^2\theta_{23}$ and the likelihood profiles as functions of $\Delta m^2_{32}$ and $\sin^2\theta_{23}$. The best fit point at $\Delta m^2_{32}=2.40^{+0.08}_{-0.09} \times 10^{-3}$\,eV$^2$ and $\sin^2\theta_{23} = 0.43^{+0.20}_{-0.04}$ weakly favors non-maximal mixing at 0.91\unit\,$\sigma$ and the normal hierarchy at 0.45\,$\sigma$. This measurement of $\Delta m^2_{32}$ is competitive with that measured by T2K and NO$\nu$A. Differences between the best fit values of the parameters in Fig.\ref{MINOS and MINOS+ Contours} are providing added precision to global fits based on these 90\%C.L. (1.6$\sigma$) contours.

In summary, analysis of the \numu disappearance and \nue appearance samples from the complete beam and atmospheric data sets of the MINOS and MINOS+ run phases has been presented 
and provides new and competitive constraints on the oscillation parameters $\Delta m^2_{32}$ and $\sin^2\theta_{23}$, weakly favors non-maximal mixing, and exhibits octant degeneracy.

\begin{acknowledgments}
This work was supported by the U.S. DOE; the United Kingdom STFC, part of UKRI; the U.S. NSF; the State and University of Minnesota; and Brazil's FAPESP, CNPq and CAPES. We are grateful to Fermilab and the Minnesota Department of Natural Resources and the personnel of the Soudan Laboratory. We thank the Texas Advanced Computing Center at The University of Texas at Austin for the provision of computing resources.
\end{acknowledgments}

\bibliography{main}% Produces the bibliography via BibTeX.

\end{document}